\documentclass[12pt,a4paper]{spie}  %>>> use this instead for A4 paper
\usepackage[]{graphicx}
\usepackage{times}

\title{
Potential of the FLASH FEL technology for the construction of a kW-scale light
source for the next generation lithography\footnote{~ Submitted to Journal of Micro/Nanolithography, MEMS and MOEMS (JM3)}
}

\author{
E.A.~Schneidmiller,
V.F.~Vogel,
H.~Weise,
M.V.~Yurkov
\skiplinehalf
Deutsches Elektronen-Synchrotron (DESY), 22607 Hamburg, Germany
}

\authorinfo{Send correspondence to M.Y.:
E-mail: mikhail.yurkov@desy.de, Telephone: +49 40 8998 2676}
\begin{document}
\maketitle

\begin{abstract}

The driving engine of the Free Electron Laser in Hamburg (FLASH) is an L-band
superconducting accelerator. It is designed to operate in burst mode with 800
microsecond pulse duration at a repetition rate of 10 Hz. The maximum
accelerated beam current during the macropulse is 9 mA. Our analysis shows that
the FLASH technology has great potential since it is possible to construct a
FLASH like free electron laser operating at the wavelength of 13.5 and 6.8
nanometer with an average power up to 2.6 kW. Such a source meets the physical
requirements for the light source for the next generation lithography.

\end{abstract}


\section{Introduction}

The roadmap for the development of the next generation lithography has been
formulated by industry in the middle of the 1990s (see \cite{bakshi-1,bakshi-2}
and references therein). A goal was to follow Moore's law in the reduction of a
feature size by a factor of two in two years \cite{moore-law}. During the last
two decades the progress in the feature size reduction has been provided by
reducing the wavelength of light sources (lasers) used. Now conventional lasers
reached their limit. Immersion and photoresist technologies aiming for a
feature size below the $\lambda /4 $ limit reached their boundaries as well.
That is why it was decided to base the next generation lithography (NGL) on a
completely new light source -- a plasma source producing extreme ultraviolet
light at the wavelength of 13.5 nm aiming for a feature size of 20 nm.
Microelectronic technology itself contains several challenging technologies
inside such as mask, reticle, resist, stepper, etc. Here we discuss only the
problem of the light source. 15 years of development of light sources for the
NGL allowed to make significant progress, and a 20~W source for a beta-level
EUV scanner has been delivered to industry \cite{bakshi-3}. However, it is
generally accepted that the problem of light source for high volume
manufacturing (HVM) is not solved yet.

Requirements for the source are not fixed but evolve with the evolution of all
the other technologies involved. For instance, requirement for the average
power (after intermediate focus (IF), in-band spectrum $\pm 2$\% around 13.5
nm, etendue of the source output $<$ 3 mm$^2$-sr) has been increased nearly by
an order of magnitude during the last ten years, and approaches two hundred
Watts.

During the last years we observed rapid progress in the development of VUV
and x-ray free electron lasers. The jump in the wavelength was about five
orders of magnitude, from 12 $\mu $m in 1997 down to 0.12 nm in 2009
\cite{pellegrini-sase,leutl,
design-rep-ttf,
flash-109nm,
FLASH-13nm,
flash-2010,
LCLS-lasing}. FLASH (Free
Electron Laser in Hamburg) at DESY has produced unprecedented powers of EUV radiation
at wavelengths of 13.5~nm and 6.8 nm, target wavelengths for the next
generation lithography (NGL) \cite{FLASH-13nm,flash-2010}.
Several schemes for a dedicated FEL-based radiation source for the NGL have
been proposed during last years \cite{ newnam-1991, litho-2000, anke-litho,
goldstein-fel05, flash-barselona, minehara-barselona, fel-2009, jinr-ngl-1,
jinr-ngl-2, socol-prstab}. In this paper we describe the potential application
for the NGL of an accelerator and FEL technology developed in the framework of
the FLASH project at DESY. We show that this technology allows
to construct a free electron laser operating at the wavelength of 13.5 nm
and 6.8 nm with an average power of several kilowatts. This becomes possible
due to two main factors. The first one is high average power (about 100 kW) in
the electron beam of a superconducting accelerator. The second important factor
is undulator tapering allowing significant increase of FEL efficiency above
2\%. As a result, high average radiation power of about several kilowatts can
be generated. Additional application of the electron beam energy recovery would
allow to increase the duty factor significantly such that the average radiation
power will be in the multi-ten-kW range \cite{litho-2000}. Spectral width of
the radiation (below 2\%) fits well to the requirements of the radiation source
for the Next Generation Lithography.

\section{Principle of FEL operation}

A free electron laser is a system consisting of a relativistic electron beam and
a radiation field interacting with each other while they propagate through an
undulator \cite{laser-handbook}. The undulator is a periodical magnetic structure
forcing electrons to perform transverse oscillations. When an electron beam
traverses an undulator, it emits radiation at the resonance wavelength $\lambda
= (\lambda_{\mathrm{w}}/2\gamma^{2})(1+K^{2}/2)$. Here, $\lambda_{\mathrm{w}}$
is the undulator period, $mc^{2}\gamma$ is the electron beam energy, $K =
eH_{\mathrm{w}}\lambda_{\mathrm{w}}/2\pi mc$ is the dimensionless undulator
strength parameter, $H_{\mathrm{w}}$ is the maximum on-axis magnetic field
strength of the undulator, $\gamma $ is the relativistic factor, $c$ is
the velocity of light, and $(-e)$ and $m$ are charge and mass of the electron,
respectively. The electromagnetic wave is always faster than the electrons, and
a resonant condition occurs when the radiation slips a distance $\lambda$
relative to the electrons after one undulator period. The fields produced by
the moving charges in one part of the undulator react on moving charges in
another part of the undulator. Thus, we deal with tail-head instability leading
to a growing concentration of particles wherever a small perturbation started
to occur. More details on FEL physics can be found in ref.~\cite{book}.

FEL devices can be divided in two classes: amplifiers and oscillators. FELs
based on oscillator principle are limited on the short-wavelength side to
ultraviolet wavelengths primarily because of mirror limitations. Free electron
lasing at wavelengths shorter than ultraviolet  can be achieved with a
single-pass, high-gain FEL amplifier. The FEL collective instability in the
electron beam produces an exponential growth (along the undulator) of the
modulation of the electron density on the scale of undulator radiation
wavelength. Power gain in excess $10^7 - 10^8$ may be obtained in the short
wavelength range. Conversion efficiency of the kinetic energy of electrons to
the VUV or soft x-ray radiation can be rather high, up to a few per cent.
Average power of the electron beam can be high as well, of the order of a
hundred kilowatts. As a results, it becomes possible to generate high average
power of the radiation at the level of a few kilowatts. Modern accelerator
technologies provide high efficiency of the conversion of the net electric
power to the electron beam (of about 10 per cent), thus allowing to obtain net
efficiency of FEL on the level of a fraction of per mille. As we
already mentioned application of the electron beam recovery may
significantly increase the net efficiency.

\section{Key elements of the FLASH/TESLA technology}

An FEL amplifier starting from the shot noise in the electron beam is a device
in which the electron beam produces powerful radiation during single pass
through an undulator \cite{derbenev-xray-1982,pellegrini}. This type of devices
is frequently referred as Self Amplified Spontaneous Emission FEL (SASE FEL)
\cite{boni86}. Efficient operation of SASE FELs is possible with thorough design
of all systems involved. Here we describe the main systems related to the
FLASH/TESLA technology.

The TESLA project (TeV Energy Linear Accelerator) has been developed by an
international TESLA collaboration since the beginning of the 1990s. More than
50 institutions from 12 countries were involved in the development of
superconducting accelerator technology for a high energy electron-positron
linear collider and for a driver-linac for an X-ray free electron laser (XFEL)
\cite{TESLA-CDR,TESLA-TDR}. Later the TESLA Collaboration has been transformed
into the TESLA Technology Collaboration (TTC) aiming at technical development
for superconducting accelerators, and gave birth to two FEL projects at DESY:
FLASH FLASH and the European XFEL \cite{design-rep-ttf,XFEL-TDR}.

\subsection{Main accelerator}

All FLASH/TESLA type accelerators are based on superconducting accelerator
modules.   Each module contains eight superconducting sections. Each section is
approximately 1 meter long and operates at a resonance frequency of 1.3 GHz.
The accelerating sections are cooled by liquid helium with temperature of 2
K. The accelerating gradient may be different for different sections, and may
approach 40 MV/m for the best quality sections. The spread of accelerating
gradients is defined by the production technology chain available, and there is
clear tendency for improving the quality
\cite{sections-quality-1,sections-quality-2}. Project value for average
gradient in the accelerating module for the European XFEL is 24.3 MV/m
\cite{weise-linac2010}. Dedicated technological studies are on the way aiming
at achieving an average accelerating gradient in excess of 35 MV/m
\cite{ilc-35mvm}. These developments are performed in the framework of
International Linear Collider (ILC) project.

Up to four accelerating modules are powered by one rf station consisting of
klystron and modulator \cite{weise-linac2010,rf-exfel-choroba}. Average and
peak power of the klystron are 150 kW and 10 MW, respectively. The accelerator
operates in the burst mode. For the FLASH project macropulse repetition rate is
10 Hz, macropulse duration is 1 ms, average beam load within macropulse may
reach 10 mA. The total rf pulse length for the European XFEL is 1.37 ms, of
which 0.72 ms are required for filling the cavity with rf power, and 0.65 ms
for electron beam acceleration. Macropulse repetition rate may be increased for
the price of macropulse length reduction. Since the filling time remains the
same, higher efficiency of the accelerator is achieved at a reduced repetition
rate.

\subsection{Laser driven rf gun}

The electron beam is produced in a laser driven RF gun\cite{rf-gun-xfel}. While
the linear accelerator is superconducting, the RF gun operates at room
temperature. An important figure of merit for the beam quality is
the transverse phase space of the electron beam or emittance $\epsilon $ (it is
just the product of the beam size by the angle spread) which should not be much
larger than radiation wavelength. The physical parameters related to the
quality of the injector is the normalized emittance, $\epsilon _n = \epsilon
\gamma $. Current values of the normalized emittance achieved in the rf guns
produced for FLASH and European XFEL are in the range 0.4 - 1 mm-mrad depending
on the bunch charge \cite{pitz-gun1}. These values are sufficient for effective
generation of x-ray generation.

\subsection{Beam formation system}

Effective operation of the VUV/x-ray FEL amplifier requires a high value of the
peak beam current of about a few thousand Ampers. The typical value of the
peak beam current in the rf gun is about a few tens of Ampers. The required
value of the peak beam current is achieved in the beam formation system. During
initial stage of amplification the electron bunch is accelerated in an
off-crest rf phase and gains energy chirp such that particles in the head of
the bunch have less energy. This leads to the compression of the bunch when
the electron beam passes through the bunch compressor - a dispersive element
providing different path lengths for particles with different energy.
Additional installation of rf sections operating at a multiple frequency
improves the quality of compressed bunches. Auxiliary sections operating at a
frequency of 3.9 MHz are used in FLASH and European XFEL \cite{3.9GHz}. Two and
three bunch compressor stages are implemented in the FLASH and EXFEL
accelerators providing peak currents up to 2.5 kA and 5 kA, respectively
\cite{XFEL-TDR,bc-dohlus-zagorodnov}.

\subsection{Undulators}

Undulators are widely used as insertion devices at synchrotron radiation
facilities around the world. The most popular undulators are
designed using permanent magnet technology \cite{Elleaume,Pflueger}. This
technology reached mature level and allows to build undulators with high
quality of magnetic field required for FEL applications: relative field errors
below per mille level and trajectory errors on a few micron level
\cite{LCLS-lasing,XFEL-TDR}. Different field geometries (planar or helical) can
be realized providing the possibility to generate radiation with required
polarization properties (linear or circular). Currently the length of undulator
systems at the FEL facilities constitute tens of meters, and exceeds hundred
meters for future projects (see Table~\ref{tab:undulators}).

\begin{table}[h]
\caption{Undulators for VUV/x-ray FELs
\cite{leutl,FLASH-13nm,LCLS-lasing,fermi-cdr,tt-sase12}}
\begin{tabular}{l c c c c c c c}
& Units
& FLASH
& LEUTL
& LCLS
& FERMI
& EXFEL \\
&&&&& FEL-2 & SASE1 \\
Period length       & mm & 27.3 & 33   & 32   & 50   & 40   \\
Peak magnetic field & T  & 0.82 & 1    & 1.25 & 0.85 &  1.1 \\
Total length        & m  & 27   & 21.6 & 112  & 14.4 & 175  \\
Number of modules   & \# & 6    & 9    & 33   & 6    & 35   \\
Length of one module& m  & 4.5  & 2.4  & 3.4  & 2.8  & 5    \\
\end{tabular}
\label{tab:undulators}
\end{table}

\section{FLASH facility}

The FLASH
facility is driven by a superconducting linear accelerator and
operates on the DESY site
\cite{design-rep-ttf,
flash-109nm,
FLASH-13nm,
flash-2010}.
The aim of this project is to
serve as a user facility, and gain experience for the XFEL project realization
\cite{XFEL-TDR}. The project FLASH started as relatively small test facility
accelerator with an energy of 240 MeV passed several upgrades. Currently it
consists of 1.2 GeV accelerator and free electron laser operating in the
wavelength range from 4.2 nm up to 47 nm
\cite{design-rep-ttf,
flash-109nm,
FLASH-13nm,
flash-2010}. It produces an ultimate peak power on
a few GW level, and peak brilliance exceeding by eight orders of magnitude
those values obtained from the third generation synchrotron radiation sources.
The FLASH facility is equipped with five beam lines for user experiments
\cite{tiedtke-flash}.

\begin{figure}[tb]

\includegraphics[width=1.0\textwidth]{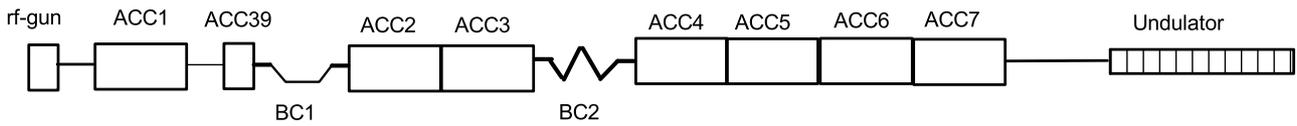}

\caption{
Schematic layout of the FLASH facility \cite{flash-2010}.
Abbreviations ACC1-ACC7, ACC39, and BC stand for accelerating module 1.3 GHz,
accelerating module 3.9 GHz, and bunch compressor,
respectively. Total length of the machine is 250~meters.
}

\label{fig:flash-2010}

\end{figure}

A schematic layout of the FLASH facility is shown in Fig.~\ref{fig:flash-2010}.
The electron beam is produced in a radio frequency gun and brought up to an
energy of up to 1200 MeV by seven accelerating modules ACC1 to ACC7 operating
at a frequency of 1.3~GHz. At intermediate energies of 150 and 450 MeV the
electron bunches are compressed in the bunch compressors BC1 and BC2. The
electron beam formation system is based on the use of linearized longitudinal
compression realized with the help of an accelerating module operating at
a frequency of 3.9~GHz. After the bunch compressor BC2, the electron beam is
accelerated to the target energy (450 to 1200 MeV), and produces powerful
coherent radiation during a single pass through the long undulator (planar,
hybrid, 12 mm fixed gap, magnetic length 27~m, period 27.3 mm, peak magnetic
field 0.48~T).

All subsystems of the superconducting linear accelerator have been optimized
for the burst mode of operation with a macropulse repetition rate of 10 Hz,
flat-top macropulse duration of 800~$\mu $s, and maximum beam loading 9~mA
within the macropulse. With these parameters the average power in the electron
beam is equal to 86 kW at the energy of 1.2~GeV. High power in the electron
beam is the main basis for the generation of high EUV radiation power.

Since 2005 FLASH operates as a user facility with high duty factor during user
runs: 7 days per week and 24 hours per day \cite{flash-2010,tiedtke-flash}. One
maintenance shift per week is foreseen for technical service and hardware
developments when it is necessary. The operation schedule includes also
dedicated runs for technical developments of accelerator and FEL subsystems.
One of the tasks which is supported by the program of International Linear
Collider is demonstration of reliable operation of the FLASH accelerator with 9
mA beam loading. There is definite progress on this way, and now we can state
that this is technically possible \cite{9ma}. The next step under discussion is
to demonstrate lasing at full beam loading. Currently maximum average radiation
power has been demonstrated during user runs with small beam load. Macropulse
repetition rate was 10 Hz, single-pulse energy was up to 0.2 - 0.3 mJ, the
number of pulses per train was up to 100 which corresponds to an average
radiation power about 0.2 - 0.3 Watt. At the moment we do not see severe
obstacles preventing lasing of full train at full beam load. However, our
observation is that the high beam load program will benefit a lot with
construction of a dedicated test facility. With FLASH being a user facility
with only a small fraction of scientific applications requesting ultimate level
of radiation power, the investigation of full power capabilities of the
FLASH/TESLA technology is unfortunately limited.

\section{European x-ray free electron laser}

The internationally organized European X-ray free electron laser (XFEL) is
under construction in Hamburg \cite{XFEL-TDR}. The construction of the linac is
a common effort of many institutes sharing the responsibility for this
superconducting linac. The overall coordination is with DESY chairing the XFEL
Accelerator Consortium \cite{weise-linac2010}. Main
contributions come from CEA Saclay/IRFU, France; CNRS LAL, Orsay, France; DESY,
Germany; INFN Milano, Italy; Soltan Institute, Poland; CIEMAT, Spain. The
project is the first large scale application of the TESLA technology developed
over the last 15 years. Superconducting accelerating cavities will be used to
accelerate the electron beam to an energy of up to 17.5 GeV.
In a start-up scenario European XFEL will be equipped with three undulator
beamlines with total magnetic length of undulators about 450 meters. The
facility will cover continuosly wavelength ranges from 0.05 nm to 60 nm.

The accelerator will consist of 100 accelerating modules, i.e. 800
superconducting accelerating structures, operating at a gradient of 24.3 MV/m.
A total of 25 RF stations will supply the necessary RF power of typically 5.2
MW per four modules. Construction work along the 3.4 km European XFEL facility
started early 2009. Start of the accelerator commissioning is scheduled for
2015. It is important to notice the high level of industrialization of the main
components of the accelerator for the European XFEL involving such companies as
Thales, Toshiba, CPI, Research Instruments and E. Zanon \cite{weise-linac2010}.

\section{Dedicated FEL source for NGL}

\begin{table}[b]
\caption{
Parameters of FLASH-NGL radiation source operating at 13.5 nm}
\medskip
\begin{tabular}{ l c c c c }
& FLASH & NGL-680 & NGL-1250 & NGL-2500 \\
Electron energy, MeV              & 680   & 680 & 1250 & 2500\\
Bunch charge, nC                  & 1    & 1    & 1    & 1 \\
Peak current, A                   & 2500 & 2500 & 2500 & 2500 \\
Normalized emittance, mm-mrad     & 1.5  & 1.5  & 1.5  & 1.5 \\
rms energy spread, MeV            & 0.5  & 0.5  & 0.5  & 0.5 \\
Macropulse duration, ms           & 0.8  & 0.8  & 0.8  & 0.8 \\
Micropulse rep. rate, MHz         & 9    & 10   & 10   & 10 \\
\# pulses in macropulse           & 7200 & 8000 & 8000 & 8000 \\
Macropulse rep. rate, Hz          & 10   & 10   & 10   & 10 \\
Undulator period, cm              & 2.73 & 2.73 & 3.7  & 5.0 \\
Undulator length, m               & 27   & 30   & 30   & 30 \\
Radiation wavelength, nm          & 13.5 & 13.5 & 13.5 & 13.5 \\
FWHM spectrum bandwidth, \%       & 0.7  & 0.7  & 0.7  & 0.7 \\
Energy in the radiation pulse, mJ & 1.4  & 8.5  & 22   & 33 \\
Peak power, GW                    & 5.6  & 34   & 88   & 130 \\
FWHM pulse duration, fs           & 250  & 250  & 250  & 250 \\
FWHM spot size, mm                & 0.17 & 0.3  & 0.2  & 0.1 \\
FWHM angular divergence, $\mu $rad& 30   & 48   & 54   & 64 \\
Average radiation power, W        & 100  & 680  & 1760 & 2640 \\
\end{tabular}
\label{table:ngl-source}
\end{table}

\begin{figure}[tb]

\includegraphics[width=0.5\textwidth]{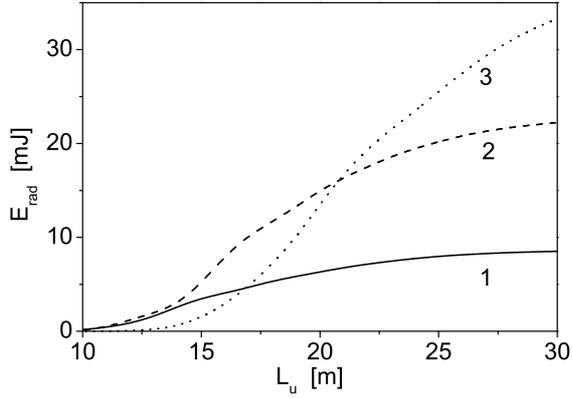}

\caption{
Energy in the radiation pulse versus undulator length. Curves 1, 2, and 3
correspond to the energy of the driving electron beam 680, 1250, and 2500 MeV,
respectively.
Radiation wavelength is 13.5 nm.
Parameters of the electron beam and FEL are given in Table
\ref{table:ngl-source}.
}

\label{fig:pz-ngl}

\end{figure}

\begin{figure}[tb]

\includegraphics[width=0.5\textwidth]{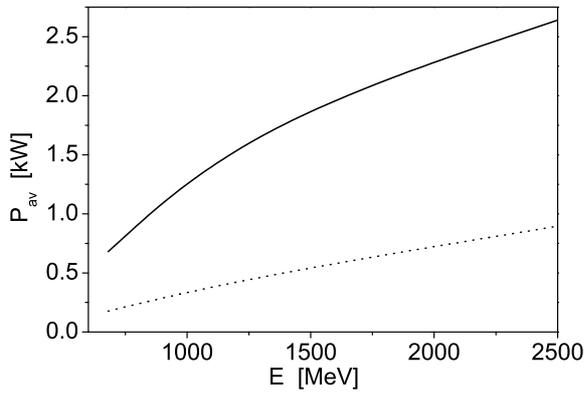}

\caption{
Average power of FEL based source for NGL as a function of the electron
beam energy. Dashed curve present FEL power in saturation, and solid curve is
FEL power at the undulator length of 30~m with optimum tapering of the
undulator parameters.
Reference parameter sets for energy 680, 1250, and 2500 MeV are
presented in Table~\ref{table:ngl-source}.
}

\label{fig:pav-30m}

\end{figure}

\begin{figure}[tb]

\includegraphics[width=0.5\textwidth]{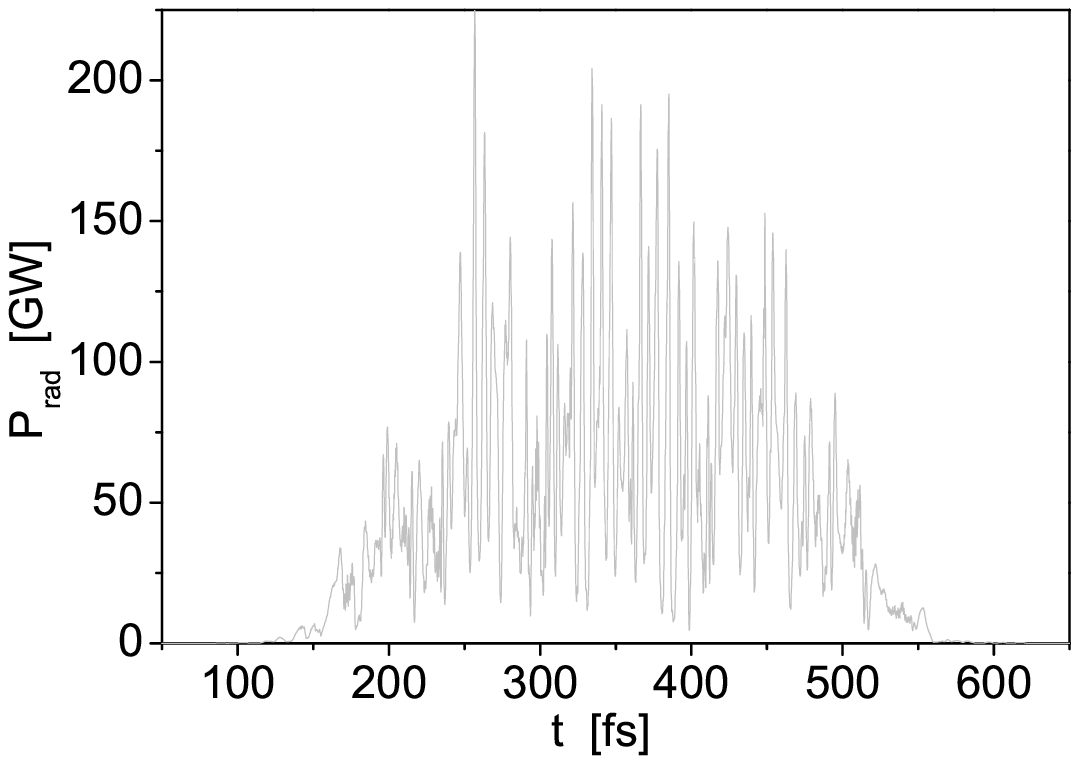}
\includegraphics[width=0.5\textwidth]{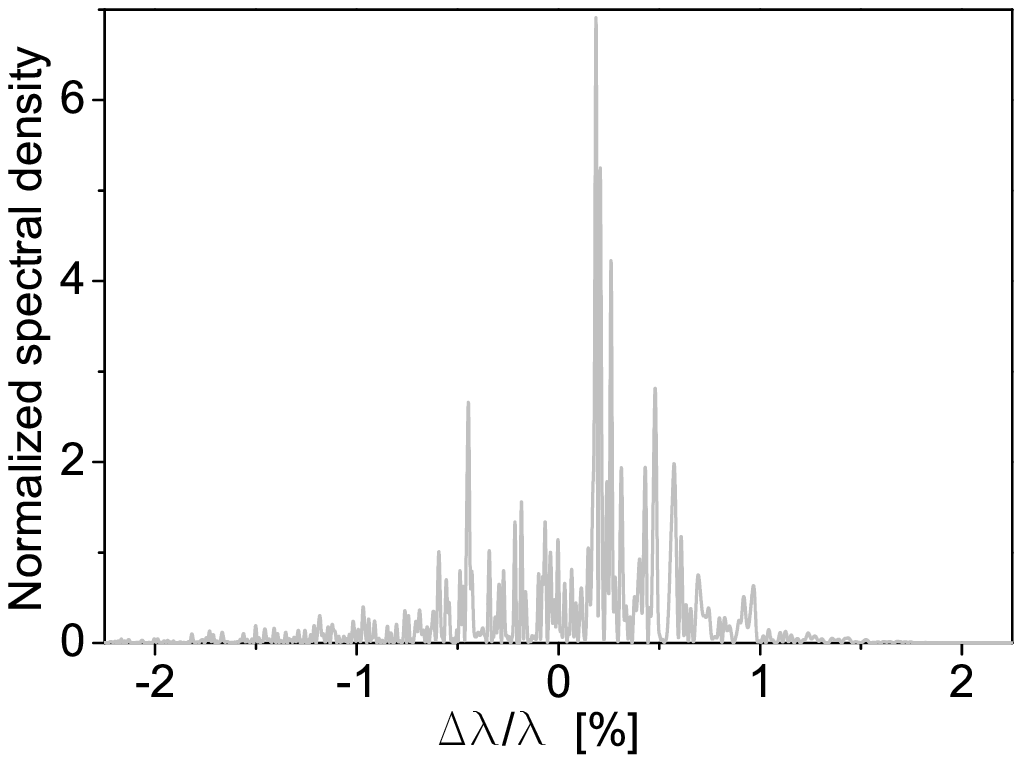}

\caption{
Temporal (left) and spectral (right) structure of the radiation pulse. FEL
parameters correspond to NGL-1250 option compiled in
Table~\ref{table:ngl-source}.
}

\label{fig:p0024029}

\end{figure}

\begin{figure}[tb]

\includegraphics[width=0.5\textwidth]{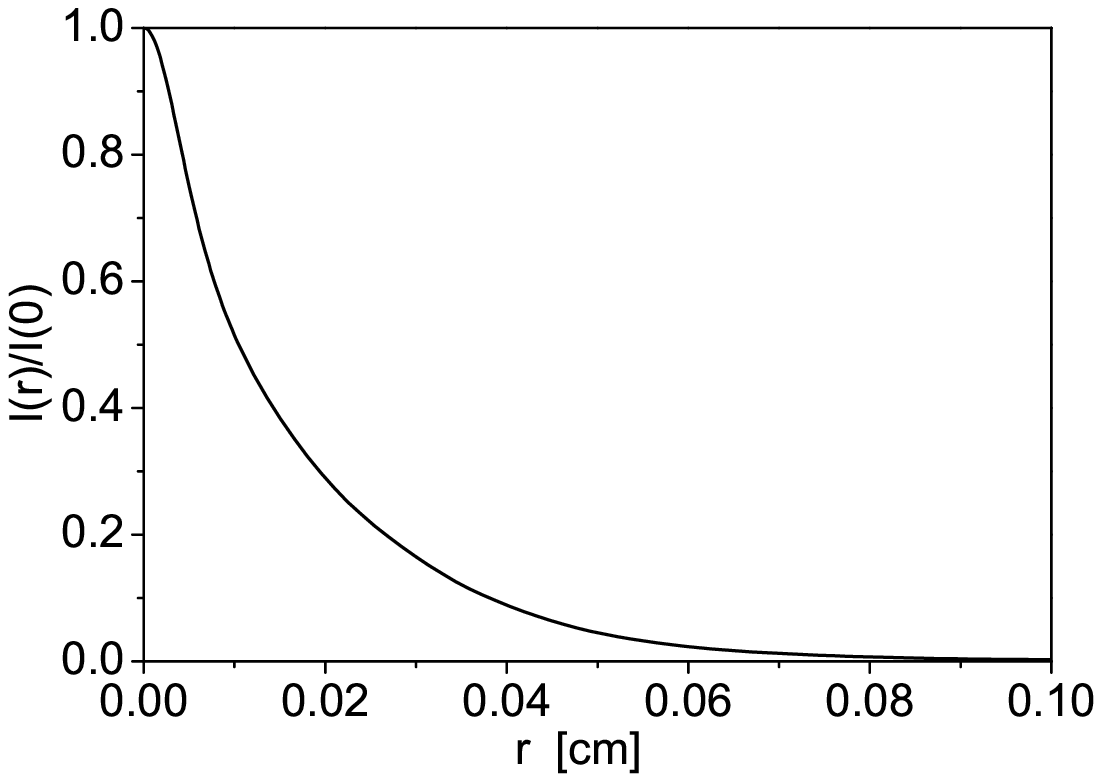}
\includegraphics[width=0.5\textwidth]{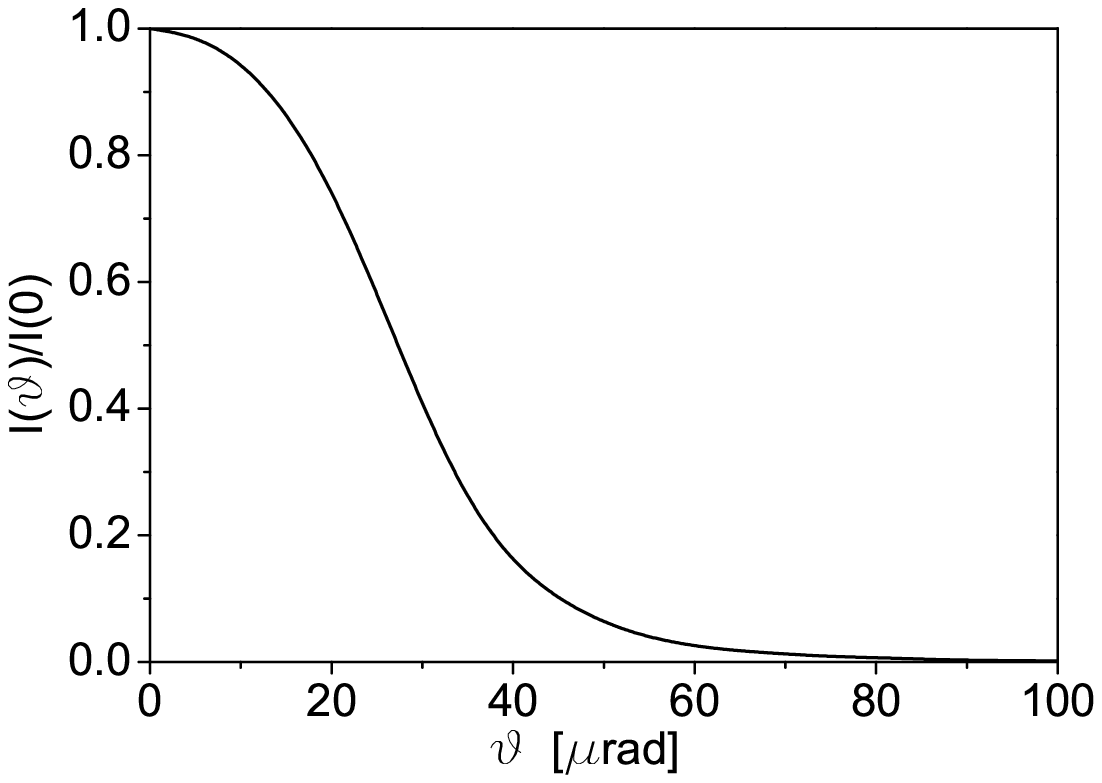}

\caption{
Distribution of the radiation intensity in the near and far
zone (left and right plot, respectively). FEL parameters correspond to NGL-1250
option compiled in Table~\ref{table:ngl-source}.
}

\label{fig:n0024029}

\end{figure}

In this section we demonstrate the potential of the FLASH/TESLA tecnology for
generation of high average power radiation. First proposals of such kind
appeared at an early stage of FLASH/TESLA technology
development\cite{litho-2000,indus-ttf,mw-fel-nim}. Presently this technology
reached mature level, and many of the project goals and milestones have been
achieved. Moreover, a large scale facility, namely European XFEL, is under
construction which is totally based on FLASH/TESLA technology. Nearly all
components for this facility are produced by industrial companies, and the next
turn is payback of FLASH/TESLA technology to edge technological areas like the
Next Generation Lithography. In the present feasibility study we do not deviate
much from the project numbers for components used in the FLASH and European
XFEL projects. The only principal extension is application of undulator
tapering to increase FEL efficiency\cite{kroll-1981} which already has been
used in early proposals of high average power FELs for industrial
applications\cite{indus-ttf,mw-fel-nim}. Recently principle of undulator
tapering has been demonstrated experimentally \cite{bnl-tapering,lcls-tapering}.

We assume operation of the linac in the burst mode with macropulse repetition
rate of 10 Hz. Macropulse duration is equal to 1 ms. Bunches with a charge of 1
nC are separated by time interval of 100 ns (10 MHz repetition rate) which
corresponds to 10 mA beam load within the macropulse. With these parameters
average power in the electron beam is equal to 100 kW at the energy of
electrons of 1 GeV.

\subsection{Operation at the wavelength of 13.5 nm}

We consider SASE FEL (single-pass FEL amplifier starting from shot noise in the
electron beam) operation. In the following we optimize the SASE FEL operation
at the wavelength of 13.5 nm. We assume parameters of the electron beam as
given in Table~\ref{table:ngl-source}. This parameter set is compiled on the
base of our experimental experience from FLASH . At an appropriate optimization
of the beam formation system we can expect the value of the normalized
emittance below 1 mm-mrad. However, a conservative value of normalized emittance
of 1.5 mm-mrad is sufficient for effective operation of a 13.5 nm FEL, and we use
this number as baseline parameter in this section. We optimize the FEL
parameters by varying the energy of the driving electron beam between 625 Mev
and 2500 MeV.

The undulator is assumed to be hybrid planar with variable gap. General feature
is that shorter gain length is achieved at a shorter undulator period and at
smaller undulator gap. In the case under study we deal with high power electron
and photon beams, and the matter of concern is to provide sufficient aperture
for the transport of both beams. We set the limit on the minimum undulator gap
to 1.2 cm like in the FLASH project.  We assume the length of the undulator
segments to be 200 cm, total magnetic length of the undulator is 30 meters, and
focusing beta function is equal to 200 cm. Calculations are performed with the
time-dependent simulation code FAST\cite{fast}.

Results of the simulations are shown in Fig.~\ref{fig:pz-ngl}. We obtain that
within the given parameter space undulator tapering is a very powerful tool for
increasing the FEL efficiency, roughly by about a factor of five with respect
to the saturation efficiency. The results obtained are very impressive: for a
chosen electron energy of 2500 MeV, the average radiation power exceeds 2.5 kW
(see Fig.~\ref{fig:pav-30m}). At a reduced electron energy (1250 MeV) we obtain
an average radiation power of 1.7 kW. The FEL efficiency is about 1.8\% in this
case. In conclusion we present the main characteristics of the radiation pulse
in Figs.~\ref{fig:p0024029} and \ref{fig:n0024029}: temporal and spectral
properties, intensity distribution at the undulator exit, and angular
distribution in the far zone. Note that phase volume of the radiation is nearly
diffraction limited, i.e. about radiation wavelength. Spectrum width is also
small, and all radiation power goes within "in-band" requirements for the NGL
radiation source.

\subsection{Operation at the wavelength of 6.8 nm}

\begin{table}[b]
\caption{Parameters of FLASH-NGL radiation source operating at 6.8 nm}
\medskip
\begin{tabular}{ l c c c c }
Electron energy, MeV               & 1250 & 1250  & 2500 & 2500 \\
Bunch charge, nC                   & 1    & 1     & 1    & 1        \\
Peak current, A                    & 2500 & 2500  & 2500 & 2500     \\
Normalized emittance, mm-mrad      & 1.5  & 1     & 1.5  & 1        \\
rms energy spread, MeV             & 0.5  & 0.5   & 0.5  & 0.5      \\
Macropulse duration, ms            & 0.8  & 0.8   & 0.8  & 0.8      \\
Micropulse rep. rate, MHz          & 10   & 10    & 10   & 10       \\
\# pulses in macropulse            & 8000 & 8000  & 8000 & 8000     \\
Macropulse rep. rate, Hz           & 10   & 10    & 10   & 10       \\
Undulator period, cm               & 3.7  & 3.7   & 5.0  & 5.0      \\
Undulator length, m                & 30   & 30    & 30   & 30       \\
Radiation wavelength, nm           & 6.8  & 6.8   & 6.8  & 6.8      \\
FWHM spectrum bandwidth, \%        & 0.7  & 0.7   & 0.7  & 0.7        \\
Energy in the radiation pulse, mJ  & 11   & 16    & 20   & 28       \\
Peak power, GW                     & 44   & 64    & 80   & 110      \\
FWHM pulse duration, fs            & 250  & 250   & 250  & 250      \\
FWHM spot size, mm                 & 0.11 & 0.12  & 0.09  & 0.08      \\
FWHM angular divergence, $\mu $rad & 35   & 38    & 41   & 41       \\
Average radiation power, W         & 880  & 1280  & 1600 & 2240     \\
\end{tabular}
\label{table:ngl-source-68}
\end{table}

\begin{figure}[tb]

\includegraphics[width=0.5\textwidth]{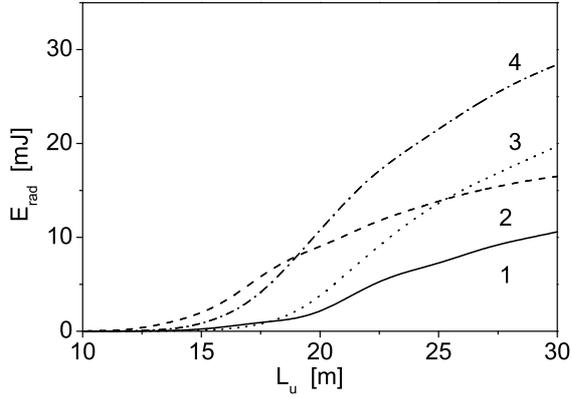}

\caption{
Energy in the radiation pulse versus undulator length. Curves 1 and 2
refer to the case of an electron energy of 1250 Mev and normalized emittance
1.5 mm-mrad and 1 mm-mrad, respectively.
Curves 3 and 4
refer to the case of an electron energy of 2500 Mev and normalized emittance
1.5 mm-mrad and 1 mm-mrad, respectively.
Radiation wavelength is 6.8 nm.
Parameters of the electron beam and the FEL are given in Table
\ref{table:ngl-source-68}.
}
\label{fig:pz-ngl-68}

\end{figure}

\begin{figure}[tb]

\includegraphics[width=0.5\textwidth]{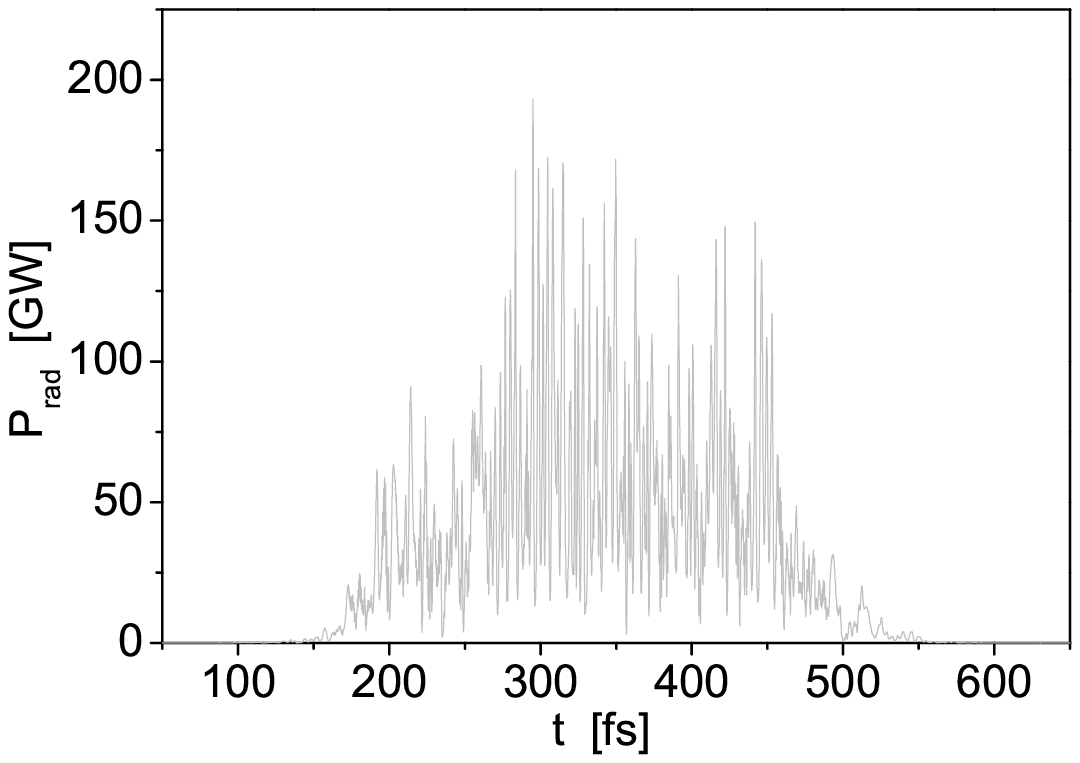}
\includegraphics[width=0.5\textwidth]{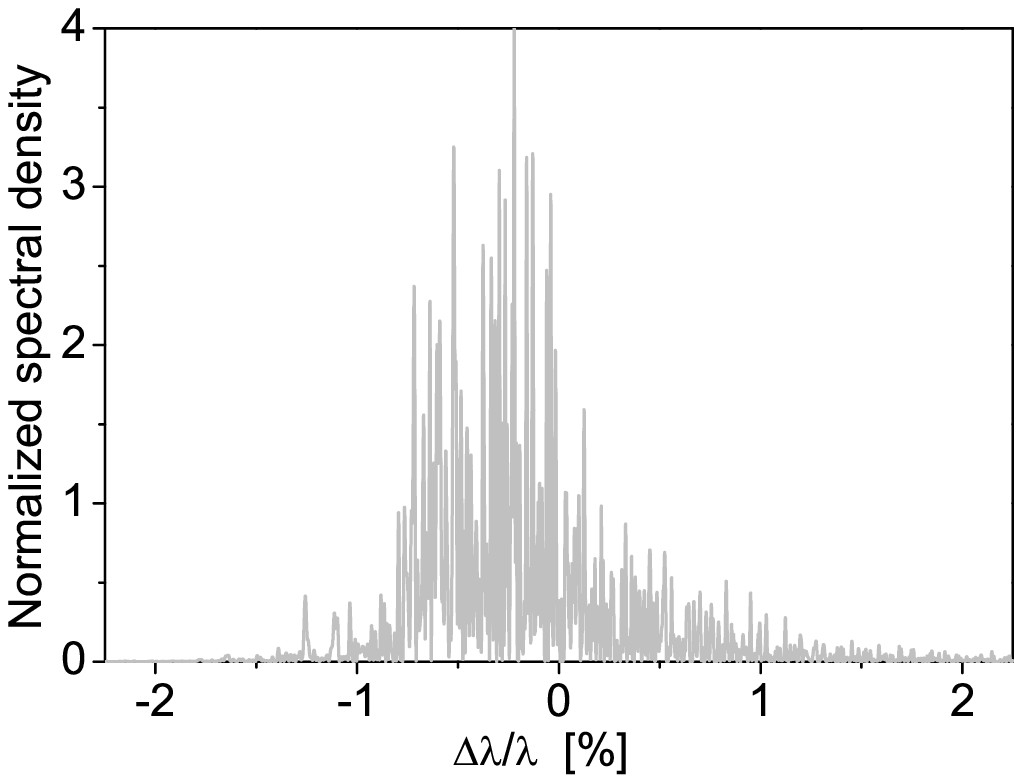}

\caption{
Temporal (left) and spectral (right) structure of the radiation pulse.
The FEL operates at a wavelength 6.8 nm. The electron energy is
equal to 1250 MeV. Normalized emittance is 1 mm-mrad (see
Table~\ref{table:ngl-source-68}).
}

\label{fig:p0071027}

\end{figure}

\begin{figure}[tb]

\includegraphics[width=0.5\textwidth]{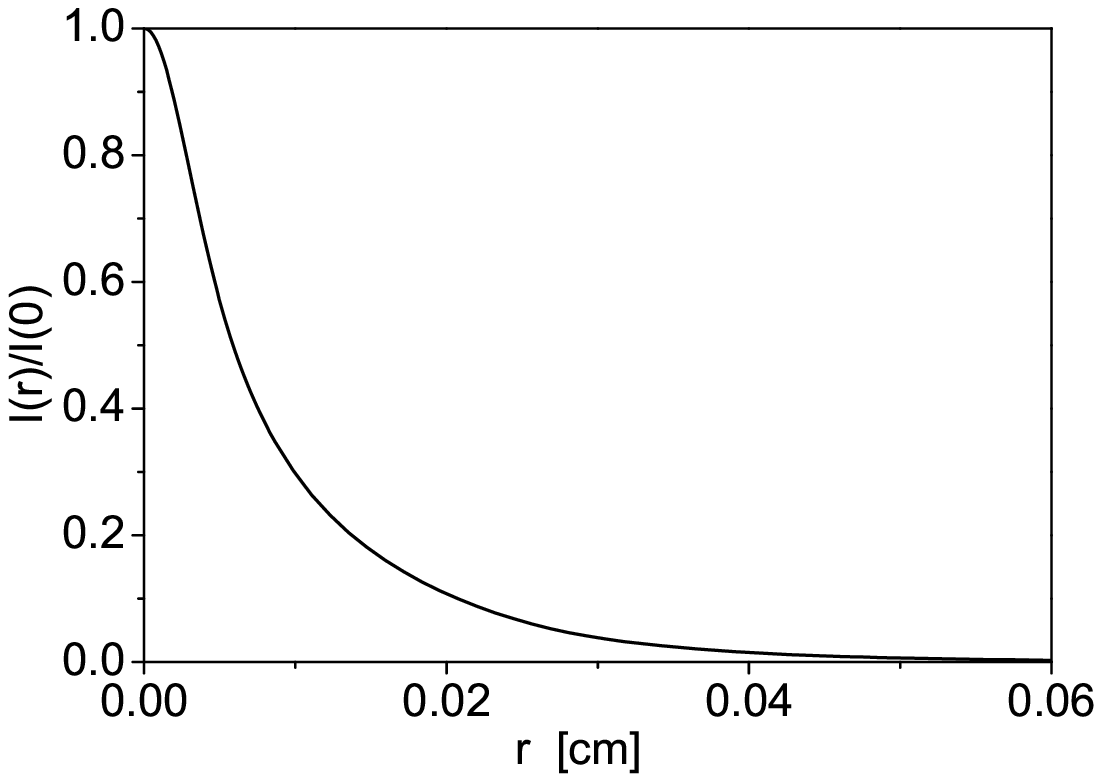}
\includegraphics[width=0.5\textwidth]{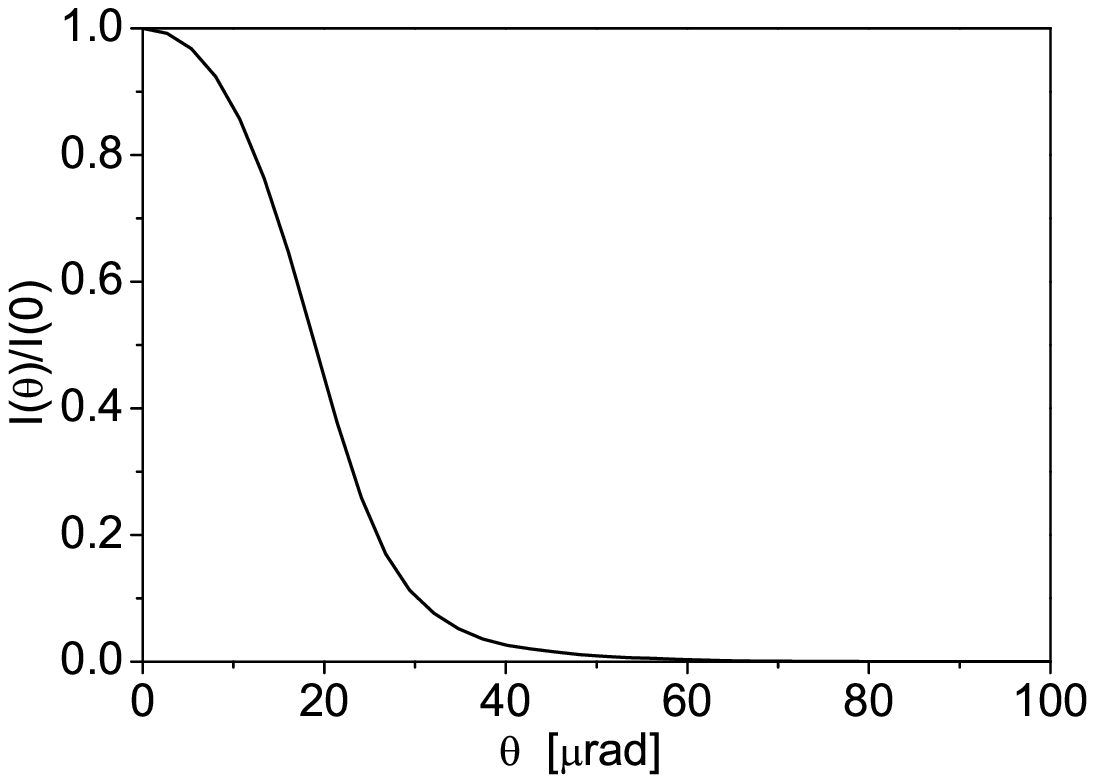}

\caption{
Distribution of the radiation intensity in the near and far zone (left and
right plot, respectively). FEL operates at the wavelength 6.8
nm. Energy of electrons is equal to 1250 MeV. Normalized emittance is 1 mm-mrad
(see Table~\ref{table:ngl-source-68}).
}

\label{fig:n0071027}

\end{figure}

The FEL is tunable radiation source, and there is a simple deal to go over to
different wavelength, for instance, to the next target wavelength of 6.8 nm
discussed in the NGL community. We restrict our study electron with energies of
1250 MeV and 2500 MeV, and use undulators optimized for
generation of 13.5 nm radiation (see previous section). Tuning of the
wavelength is performed by the increase of the undulator gap from 12 mm to 18
mm. Magnetic field changes from 1 T to 0.7 T. Evolution of the energy in the
radiation pulse is shown in Fig.~\ref{fig:pz-ngl-68}. Comparison of the pulse
energies at 6.8 nm and 13.5 nm shows reduction of the pulse energy when going
to shorter wavelength. Partially this reduction relates to nonoptimal
undulator: period length has been optimized for 13.5 nm. Another factor is
reduction of the FEL gain and FEL efficiency due to shorterning of the
wavelength. FEL elefficiency can be increased with the reduction of the electron
beam emittance. Curves 2 and 4 in Fig.~\ref{fig:pz-ngl-68} show gain curves of
the SASE FEL driven by the electron beam with normalized emittance of 1 mm-mrad.
Output power in the case becomes to be comparabable with the case of the
wavelength 13.5 nm and $\epsilon _n = 1.5$ mm-mrad. Recent developments of the
laser driven rf gun demonstrated feasibility for generation of the beams
with rms normalized emittance well below 1 mm-mrad \cite{pitz-gun1}. Thus, we
can conclude that it is technically feasible to produce 2 kW level of the
average radiation power at the wavelength of 6.8 nm. It is important that the
same hardware (accelerator and undulator) is used for production of the
radiation with wavelength of 13.5 nm and 6.8 nm. Key parameters of the
radiation: temporal and spectral structure of the radiation pulse, intensity
distribution of the radiation (see Figs.~\ref{fig:p0071027} and
\ref{fig:n0071027}) do not differ much from characteristics of the source tuned
to 13.5 nm (see Figs.~\ref{fig:p0024029} and \ref{fig:n0024029}).

\section{Discussion}

Up to now two technologies for the light source are developed by industry:
laser produced plasma (LPP), and discharge produced plasma (DPP)
\cite{bakshi-1,bakshi-2}. Serious obstacle on the way to a HVM source is the
small efficiency of these sources which requires deposition of huge power in a
small volume. Another problem is the mitigation of the plasma debris required
for the protection of EUV optics. An FEL based
radiation source has evident advantages. The process of light generation takes
place in vacuum, and there is no problem to utilize the spent electron beam
(remove unused power). The problem of debris mitigation does not exists at all.
There is no collector problem since the radiation is produced in the
diffraction limited volume, and there is no problem with the transport of the
radiation to the exposure tool.

\begin{figure}[tb]

\includegraphics[width=0.7\textwidth]{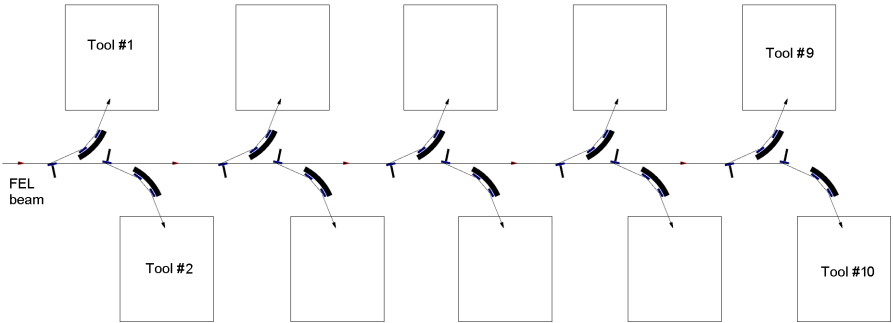}

\caption{
Integration of an FEL-based NGL source in the factory using an approach of
"single source" - "multiple tools". }

\label{fig:multi-tool}

\end{figure}

Laser and plasma radiation sources may operate only at a limited set of discrete
wavelengths. Each dedicated source operates at a fixed wavelength only. With
a SASE FEL driven by 1 - 2 GeV electron beam, the available wavelength range
can span from about 100 nm down to a few nanometers. This feature can be
helpful for instance, for optimization of a photoresist. It is important that
the same hardware (accelerator and undulator) can be used for production of
tunable powerful radiation as we demonstrated in our paper. Average power of
this "clean" EUV radiation is in the range of a few kilowatts, an order of
magnitude above present requirements for a HVM source. Thus, one FEL set up can
replace a dozen of plasma sources. Currently microelectronic industry uses an
approach of a single source for a single tool (stepper). In the case of using a
powerful FEL as the source we can modify this approach to a single source for
multiple tools. We can discuss, for instance, transport of the photon beam to
different tools allowing their quasi-simultaneous operation
\cite{flash-barselona,photon-ring,TESLA-FEL-2004-02} (see
Fig.~\ref{fig:multi-tool}). Another option which is under discussion is
generation of the radiation in several undulators quasi-simultaneously
\cite{minehara-barselona}.

The parameter space considered in this conceptual study does not deviate much
from project parameters of FLASH/TESLA technology. Main idea was
the demonstration of the capabilities of this technology to produce multi-kW
level of output radiation power and show that properties of the radiation meet
requirements of the Next Generation Lithography. It is important that
the described device can be constructed now without additional R\&D using
components developed for FLASH and European XFEL. Definitely, there is a room
for further optimization of the source. This can be essential subject for joint
collaboration between R\&D departments of microelectronic industry and
developers of accelerator and FEL techniques. For instance, it is rather easy
to increase average output power of the source by means of increasing the
length of macropulse and corresponding reduction of the repetition rate.
Alternatively, macropulse repetition rate can be increased for the price of
reduction of the output power. This kind of considerations has been already
analyzed in the framework of the European XFEL project \cite{XFEL-TDR}.

Within long term of FLASH/TESLA technology development we can discuss
extensions towards increase of duty factor up to cw mode of operation. This
activity is also performed in the framework of TESLA Technical Collaboration.
One of the problems to be solved is that of high duty cycle or cw injectors
generating low emittance beams \cite{hw3,hw4}. The necessary replacement of
klystrons by Inductive Output Tubes (IOT) seems to be possible \cite{hw5}. When
the injector technology becomes available, the FEL based radiation source can
operate in cw mode. Application of energy recovery will allow to go over to
higher output radiation powers.

\section*{Acknowledgements}

We thank R. Brinkmann for interest in this work and support. We thank O.
Gushchin, E. Saldin, and G. Shirkov for their interest in this work. We thank E.
Syresin for useful discussions. During the EUVL Symposium in Barcelona and
several meetings later on we had the possibility for stimulating discussions
with our colleagues from the FEL community and from industry: V. Banine, U.
Dinger, A. Endo, A. Goehnermeier, M. Goldstein, M. Kemp, E. Minehara, R. Moors,
J.H. Peters, E. Sohmen, U. Stamm, X.J. Wang, and P. Willemse. We also thank V.
Bakshi and A. Endo for an invitation to discuss the potential of FELs for the
next generation lithography.


\begin{thebibliography}{99}

\bibitem{bakshi-1}
Vivek Bakshi (Editor),
EUV Sources for Lithography, SPIE Press Monograph,  Vol. PM149, 2006.

\bibitem{bakshi-2}
Vivek Bakshi (Editor),
EUV Lithography, SPIE Press Book, Vol. PM178, 2008.

\bibitem{moore-law}
Gordon E. Moore, Electronics, Volume 38, Number 8, April 19, 1965.

\bibitem{bakshi-3}
Vivek Bakshi,
EUV Sources Come Back as Top EUV Lithography Concern,
Semiconductor International, July, 2009. http://www.semiconductor.net.

http://www.euvlitho.com/2011/2011%20SPIE%20AL%20Summary.pdf

\bibitem{pellegrini-sase}
M. Hogan et al.,
Phys. Rev. Lett. 81, p. 4867, 1998.

\bibitem{leutl}
S.V. Milton et al.,
Science, 292, p. 2037, 2001.

\bibitem{design-rep-ttf}
T.~{\AA}berg, et al., A VUV FEL at the TESLA Test Facility at DESY,
Conceptual Design Report,  DESY Print TESLA-FEL 95-03 , May 1995.

\bibitem{flash-109nm}
J.~Andruszkow et al.,
Phys. Rev. Lett. 85, p.3825, 2000.

\bibitem{FLASH-13nm}
W. Ackermann et al.,
Nature Photonics, 1, p. 336, 2007.

\bibitem{flash-2010}
S. Schreiber et al.,
Proc. FEL2010, Malmo, Sweden, 2010, TUOBI2.
http://accelconf.web.cern.ch/AccelConf/FEL2010/papers/tuobi2.pdf.

\bibitem{LCLS-lasing}
P. Emma et al.,
Nature Photonics, 4, p. 641, 2010.

\bibitem{newnam-1991}
Brian E. Newnam,
Proc. SPIE, Vol. 1343, p. 214, 1991.

\bibitem{litho-2000}
C.~Pagani et al.,
Nucl. Instrum. and Methods A 463, p.9, 2001.

\bibitem{anke-litho}
G. Dattoli et al.,
Nucl. Instrum. and Methods A 474, p.259, 2001.

\bibitem{goldstein-fel05}
M. Goldstein et al.,
Proc. 27th FEL Conference, Stanford, p. 422, 2005.

\bibitem{flash-barselona}
E.L. Saldin, E.A. Schneidmiller, H. Weise, and M.V. Yurkov,
Free electron laser as a potential source for EUV lithography,
talk at the EUVL Source Workshop, Barselona, 2006.

\bibitem{minehara-barselona}
E. Minehara et al.,
A multi-kW EUV light source driven by energy-recovery linac,
talk at the EUVL Source Workshop, Barselona, 2006.

\bibitem{fel-2009}
E.L. Saldin, E.A. Schneidmiller, V.F. Vogel, H. Weise and M.V. Yurkov,
Proc. FEL2009, Liverpool, UK, MOPC54.
http://accelconf.web.cern.ch/AccelConf/FEL2009/papers/mopc54.pdf.

\bibitem{jinr-ngl-1}
E. Syresin et al.,
Proc. RuPAC-2010, Protvino, Russia, 2010. \newline
http://accelconf.web.cern.ch/AccelConf/r10/papers/wepsb003.pdf.

\bibitem{jinr-ngl-2}
E.M. Syresin et al.,
Journal of Surface Investigation. X-ray, Synchrotron and Neutron Techniques,
5, p.520, 2011.

\bibitem{socol-prstab}
Y. Socol, G.N. Kulipanov, A.N. Matveenko, O.A. Shevchenko, and N.A. Vinokurov,
Phys. Rev. ST Accel. Beams 14, p.040702, 2011.

\bibitem{laser-handbook}
E.L. Saldin, E.A. Schneidmiller, and M.V. Yurkov,
Free-Electron Lasers, (in:
Springer Handbook of Lasers and Optics, Chapter 11 (Springer, New
York, 2007), pp. 814-819.

\bibitem{book}
E.L. Saldin, E.A. Schneidmiller, and M.V. Yurkov, The Physics of
Free Electron Lasers (Springer-Verlag, Berlin, 1999).

\bibitem{derbenev-xray-1982}
Ya.S. Derbenev, A.M. Kondratenko, and E.L. Saldin,
Nucl. Instrum. and Methods 193, p.415, 1982.

\bibitem{pellegrini}
J.B. Murphy and C. Pellegrini,
Nucl. Instrum. and Methods A 237, p.159, 1985.

\bibitem{boni86}
R. Bonifacio, F. Casagrande and L. De Salvo Souza,
Phys. Rev. A 33, p.2836, 1986.

\bibitem{TESLA-CDR}
Conceptual Design of 500 GeV e+e- Linear Collider with
Integrated X-ray Facility (Edited by R. Brinkmann et al.),
DESY 1997-048, ECFA 1997-182, Hamburg, May 1997.

\bibitem{TESLA-TDR}
TESLA Technical Design Report, DESY 2001-011, March 2001
\newline
http://tesla.desy.de/new\_pages/0000\_TESLA\_Project.html.

\bibitem{XFEL-TDR}
Altarelli, M. et al. (Eds.),
XFEL: The European X-Ray Free-Electron Laser, Technical Design Report.
Preprint DESY 2006-097, DESY, Hamburg, 2006 (see also http://xfel.desy.de).

\bibitem{sections-quality-1}
L. Lilje and D. Reschke,
Proc. LINAC08 Conference, Canada, THP014.
http://accelconf.web.cern.ch/AccelConf/LINAC08/papers/thp014.pdf.

\bibitem{sections-quality-2}
W. Singer et al.,
Proc. IPAC'10 Conference, Kyoto, Japan, 2010, THOARA02.
http://accelconf.web.cern.ch/AccelConf/IPAC10/papers/thoara02.pdf.

\bibitem{weise-linac2010}
H. Weise,
Proc. Linear Accelerator Conference LINAC2010, Tsukuba, Japan, MO102.
http://accelconf.web.cern.ch/AccelConf/LINAC2010/papers/mo102.pdf.

\bibitem{ilc-35mvm}
 N. Ohuchi et al.,
Proc. IPAC'10 Conference, Kyoto, Japan, 2010, WEPE008.
http://accelconf.web.cern.ch/AccelConf/IPAC10/papers/wepe008.pdf.

\bibitem{rf-exfel-choroba}
S. Choroba,
Proc. PAC07 Conference, Albuquerque, New Mexico, USA, 2007, TUXC03.
http://accelconf.web.cern.ch/AccelConf/p07/PAPERS/TUXC03.PDF.

\bibitem{rf-gun-xfel}
S. Rimjaem et al.,
Proc. EPAC08 Conference, Genoa, Italy, 2008, MOPC078.
http://accelconf.web.cern.ch/AccelConf/e08/papers/mopc078.pdf.

\bibitem{pitz-gun1}
S. Rimjaem et al.,
Proc. IPAC'10 Conference, Kyoto, Japan, 2010, TUPE011.
http://accelconf.web.cern.ch/AccelConf/IPAC10/papers/tupe011.pdf.

\bibitem{3.9GHz}
E. Vogel et al.,
Proc. IPAC'10 Conference, Kyoto, Japan, 2010, THPD003.
http://accelconf.web.cern.ch/AccelConf/IPAC10/papers/thpd003.pdf.

\bibitem{bc-dohlus-zagorodnov}
I. Zagorodnov and M. Dohlus,
Phys. Rev. ST Accel. Beams 14, p. 014403, 2011.

\bibitem{Elleaume}
H. Onuki and P. Elleaume,
Undulators, Wigglers, and Their Applications (Taylor
Francis, New York 2003).

\bibitem{Pflueger}
J. Pflueger,
Nucl. Instrum. and Methods A 445, p. 366, 2000.

\bibitem{fermi-cdr}
C.J. Bocchetta et al., FERMI\@Elettra CDR, Synchrotrone Trieste, 2007,
http://www.elettra.trieste.it/FERMI.

\bibitem{tt-sase12}
T. Tschentscher, Layout of X-Ray Systems, Europena XFEL Technical
Note XFEL.EU TN-2011-001.

\bibitem{tiedtke-flash}
K. Tiedtke et al.,
New J. Phys. 11, 023029, 2009.

\bibitem{9ma}
S. Pei et al.,
Proc. IPAC'10 Conference, Kyoto, Japan, 2010, TUPEA059.
http://accelconf.web.cern.ch/AccelConf/IPAC10/papers/tupea059.pdf.

\bibitem{indus-ttf}
C.~Pagani,
E.L.~Saldin,
E.A.~Schneidmiller
and M.V.~Yurkov,
Nucl. Instrum. and Methods A 423, p. 190, 1999.

\bibitem{mw-fel-nim}
C.~Pagani,
E.L.~Saldin,
E.A.~Schneidmiller
and M.V.~Yurkov,
Nucl. Instrum. and Methods A 455, p.733, 2000).

\bibitem{kroll-1981}
N. Kroll, P. Morton, M. Rosenbluth,
IEEE J. Quant. Electr. 17, p.1436, 1981.

\bibitem{bnl-tapering}
Y. Hidaka et al.,
Proc. 2001 Part. Acc. Conference, New York, USA, 2011, THP148,
http://www.c-ad.bnl.gov/pac2011/proceedings/papers/thp148.pdf.

\bibitem{lcls-tapering}
D. Ratner et al.,
Proc. FEL2009, Liverpool, UK, TUOA03. \newline
http://accelconf.web.cern.ch/AccelConf/FEL2009/papers/tuoa03.pdf.

\bibitem{fast}
E.L.~Saldin, E.A.~Schneidmiller, and M.V.~Yurkov,
Nucl. Instrum. and Methods A 429, p.233, 1999.

\bibitem{photon-ring}
E.L.~Saldin, E.A.~Schneidmiller, and M.V.~Yurkov,
Nucl. Instrum. and Methods A 507, p.510, 2003.

\bibitem{TESLA-FEL-2004-02}
E.L.~Saldin, E.A.~Schneidmiller, and M.V.~Yurkov,
The Potential for the Development of the X-Ray Free Electron Laser,
DESY Print TESLA-FEL 2004-02, Hamburg, 2004.

\bibitem{hw3}
A. Arnold et al.,
14th SRF Conference, Berlin, p.20, 2009.

\bibitem{hw4}
T. Kamps et al.,
14th SRF Conference, Berlin, p.164, 2009.

\bibitem{hw5}
Y. Li,
EUROFEL-Report-2007-DS5-074.

\end{thebibliography}
\end{document}